\documentclass[10pt,letterpaper]{article}
\usepackage{opex3}

\begin{document}

\title{A two-state Raman coupler for coherent atom optics}

\author{J. E. Debs, D. D\"{o}ring, N. P. Robins, C. Figl, P. A. Altin, and J. D. Close}

\address{Australian Centre for Quantum Atom Optics, Department of Quantum Science\\ The Australian National University, Canberra, 0200, Australia}

\email{john.debs@anu.edu.au}



\begin{abstract}
We present results on a Raman laser-system that resonantly drives a closed two-photon transition between two levels in different hyperfine ground states of $^{87}$Rb. The coupler is based on a novel optical design for producing two phase-coherent optical beams to drive a Raman transition. Operated as an outcoupler, it produces an atom laser in a single internal atomic state, with the lower divergence and  increased brightness typical of a Raman outcoupler. Due to the optical nature of the outcoupling, the two-state outcoupler is an ideal candidate for transferring photon correlations onto atom-laser beams. As our laser system couples just two hyperfine ground states, it has also been used as an internal state beamsplitter, taking the next major step towards free space Ramsey interferometry with an atom laser. 
\end{abstract}



\section{Introduction}

In recent experimental work, the correlation function $g^{(2)}(\tau)$ of an atom laser was measured using a high finesse optical cavity to count single atoms \cite{Ottl:2005aa}. These correlation measurements have extended the analogy between optical and atom lasers into the quantum regime, and pave the way to observing quadrature squeezing and entanglement in an atomic beam. Firstly, however, two criteria should be met: (i) The atom laser beam should ideally be composed of a single state, as multi-state beams, such as those produced by radio frequency (RF) or Raman coupling between Zeeman states, add classical noise and can contribute to false atom counts, and (ii), a method for producing correlations in the atomic beam that is compatible with (i) must be found. In this paper, we demonstrate an atom-laser system which meets both these criteria, based on optical Raman transitions between the hyperfine ground states of $^{87}$Rb.

\begin{figure}[b]
\begin{center}
\includegraphics{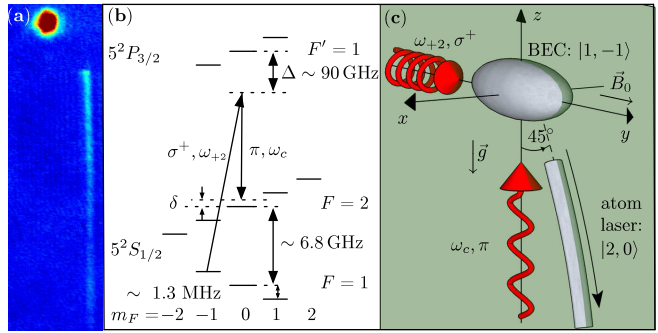}
\caption{\label{AL} (a) Typical atom laser absorption image, (b) energy level scheme for outcoupling (not to scale), and (c) Raman beam geometry.  Only the $F'=1$ manifold of the $5^2P_{3/2}$ excited state of $^{87}$Rb is shown. $\delta$ is the two-photon detuning.  The bias field, $\vec{B_0}$, is directed along the long axis of the condensate ($y$ in (c)), and defines the quantization axis. We direct the modulated Raman beam along this direction, with circular polarization to drive absorption of a $\sigma^{+}$ photon from the $\omega_{+2}$ sideband. The unmodulated Raman beam propagates upwards and is linearly polarized along $\vec{B_0}$ to drive emission of a $\pi$ photon. The two-photon transition gives the outcoupled atoms a net momentum kick of $\sqrt{2}\hbar k$ producing a beam in the continuous regime.}
\end{center}
\end{figure}

The coherence properties of Bose-Einstein condensates (BECs) led to the first demonstration of an ``atom laser" \cite{Mewes:1997aa}; a coherent beam of atoms which can be viewed as a macroscopic matter wave, in direct analogy with an optical laser. Atom laser outcouplers for magnetically trapped BECs operate by transferring atoms from trapped to untrapped Zeeman states. This is accomplished using either optical two-photon (Raman) transitions \cite{Hagley:1999aa,Jeppesen:2008aa,Robins:2006aa}, or non-optical (RF or microwave) single-photon transitions \cite{Bloch:1999aa,Dugue:2007aa,Kohl:2005aa,Le-Coq:2001aa,Riou:2006aa}, both of which have been extensively studied. RF outcoupling between Zeeman levels of a given hyperfine state is by far the most common method used. However, for typical magnetic trap bias fields, only the first order Zeeman shift is substantial and adjacent Zeeman levels are symmetrically split, resulting in coupling to all $2F+1$ sublevels, where $F$ is the total angular momentum of the hyperfine atomic state. This produces a multi-state atom laser beam that is not ideal for applications involving the measurement of quantum statistics, and unsuitable for coherent atom interferometry and precision measurement. Microwave outcoupling can be used to produce a single-state atom laser beam by coupling two Zeeman levels of different hyperfine states, which ensures that no other transitions are resonant  \cite{Ottl:2005aa,Ottl:2006aa}. However, interactions between outcoupled and trapped atoms leads to beams with complex spatial modes and a high divergence when compared with the spatial Heisenberg limit \cite{Jeppesen:2008aa,Kohl:2005aa,Le-Coq:2001aa,Riou:2006aa}. This can be avoided by outcoupling from the bottom of the condensate, which minimises the interaction time \cite{Bourdel:2006aa,Ottl:2005aa}, but also limits the flux. Outcoupling from the centre of the condensate maximises the atomic flux, and optical Raman transitions have the advantage of transferring momentum, up to a maximum of two photon recoils, to the outcoupled atoms, reducing the interaction time between trapped and outcoupled atoms. High brightness atom lasers with a gaussian spatial mode can be produced with this technique; such beams approach the Heisenberg limit \cite{Jeppesen:2008aa}. Another advantage of optical-based outcoupling over both RF and microwave outcoupling is the potential to produce correlations in an atom laser beam (or between two atom-laser beams) via quantum state transfer of a squeezed optical state, based on the system proposed by Haine \textit{et al.} \cite{Haine:2005aa,Haine:2006aa}. However, to date, Raman outcouplers have only operated by coupling Zeeman levels in a single hyperfine state \cite{superradiance}, leading to multi-state beams.

We demonstrate operation of a Raman coupler based on a novel optical design in which both beams are sourced from a single external cavity diode laser (ECDL). The coupler operates resonantly between only two Zeeman levels, one in each of the hyperfine ground state of $^{87}$Rb, and is capable of driving $\Delta m_{\textrm{F}}=\pm1,0$ transitions. The $\sim$ 6.8\,GHz splitting of the hyperfine levels results in a two-state system for a large one-photon detuning. Used as an outcoupler, this produces an atom laser in a single quantum state with all the aforementioned advantages of a Raman outcoupler, and we present results focussed primarily on atom laser production. Using $\Delta m_{\textrm{F}}=0$ transitions, our laser-system can also be operated as a hyperfine state atomic beamsplitter \cite{Kasevich:1991aa}. All these properties make it a versatile tool for coherent atom interferometry, precision measurement, and experiments investigating entanglement of massive particles.

\section{Method}

\begin{figure}
\begin{center}
\includegraphics[scale=0.2]{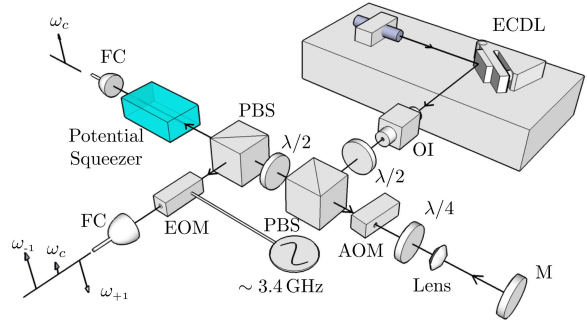}
\caption{\label{setup} Optical setup for the production of two Raman laser beams, separated in frequency by $\sim$ 6.8\,GHz. ECDL: external cavity diode laser, OI: optical isolator, $\lambda/2$: half-wave plate, $\lambda/4$: quarter-wave plate, PBS: polarizing beamsplitter cube, AOM: acousto-optic modulator used for intensity control and fast shuttering, M: mirror, FC: fiber coupler, and EOM: electro-optic modulator driven by a $\sim$ 3.4\,GHz sine wave. A phasor diagram of the frequency components in each of the two Raman beams is included after each fiber coupler. Note that higher order sidebands are present in the modulated beam, but only first order are displayed in the diagram. A potential optical squeezer is included on one Raman beam to emphasise the possibility of using the setup to transfer squeezed photon statistics to the atom laser.}
\end{center}
\end{figure}

To test and characterise the outcoupler, we prepare BECs of $^{87}$Rb in the $|F=1,m_F=-1\rangle$ state of typically $2.5\times10^5$ atoms with no discernible thermal fraction. Our Ioffe-Pritchard magnetic trap has trapping frequencies of $\omega_y= 12$ Hz and $\omega_x,\omega_z=128$ Hz, a field gradient of 200\,G/cm, and a stable bias field of 2\,G at the trap minimum allowing highly reproducible atom laser production. A typical (destructive) absorption image is given in Fig.\,\ref{AL}(a). 5\,ms of weak Raman outcoupling is applied using the three-level coupling scheme of Fig.\,\ref{AL}(b) with the Raman beam geometry shown in Fig.\,\ref{AL}(c). The magnetic trap is then suddenly switched off allowing the cloud and atom-laser beam to ballistically expand for 15\,ms. The atoms are then  illuminated by a 100\,$\mu$s pulse of resonant light, and imaged onto a CCD camera. The beam is displaced to the side, and away from the BEC for this expansion time, due to the momentum transfer from the Raman transition. It has a low divergence typical of a Raman outcoupler, and only contains atoms in a single internal state.

Production of the two Raman laser beams is schematically represented in Fig.\,\ref{setup}.  Laser light is sourced from an ECDL, red-detuned by $\Delta \simeq$ 90 GHz from one-photon resonance in order to suppress spontaneous emission. In this limit, and for a three level-system (Fig.\,\ref{AL}(b)), the coupling strength for a Raman transition is characterized by the (two-photon) Rabi frequency, $\Omega_{\textrm{\footnotesize{Rabi}}}^{(2)}=\Omega_1\Omega_2/4\Delta$, where  $\Omega_1 \propto |\vec{E}_{0,1}|$ and $\Omega_2 \propto |\vec{E}_{0,2}|$ are the one-photon Rabi frequencies for each light field with amplitudes $|\vec{E}_{0,1}|$ and $|\vec{E}_{0,2}|$ respectively. Therefore, $\Omega_{\textrm{\footnotesize{Rabi}}}^{(2)} \propto |\vec{E}_{0,1}||\vec{E}_{0,2}|/\Delta$. The light is first sent through an acousto-optic modulator in a double pass configuration to allow for intensity control, and fast shuttering on the order of 100 ns, before being split equally at a polarizing beamsplitter. One beam is sent through a phase modulator, driven at $\omega_m \simeq$ 3.4\,GHz by a Rhode \& Schwarz SMR-20 microwave generator. A phase modulated source results in sidebands according to the relation $\vec{E}=\vec{E}_0e^{i(\omega_ct+\phi/2 \cos(\omega_mt))}=\vec{E_0}e^{i\omega_ct}\sum_{n=-\infty}^{\infty}i^nJ_n(\phi/2)e^{in\omega_mt}$, where $J_n(\phi/2)$ is the $n^{th}$ order Bessel function of the first kind and represents the amplitude of the $n^{th}$ order sideband. Every pair of frequencies seperated by $2\omega_m \simeq$ 6.8\,GHz can drive the Raman transtion and contribute to the coupling strength. However, the different pairs contribute destructively for phase modulated light and $\Omega_{\textrm{\footnotesize{Rabi}}}^{(2)} \propto \sum_{n=-\infty}^{\infty}J_n(\phi)J_{n-2}(\phi/2)/\Delta$ goes to zero \cite{Lee:2003aa}. It is therefore necessary to modulate only \emph{one} Raman beam resulting in a coupling strength $\propto |\vec{E}_{0,\omega_c}|J_{\pm 2}(\phi)/\Delta$, due to one of the second order sidebands, $\omega_{\pm2}$, at $\omega_c\pm 2\omega_m$ beating with the carrier. Indeed, we have not observed Raman transitions when using phase modulated light for both Raman beams. We monitor the sidebands in the modulated beam using a commercial confocal cavity with a free spectral range of 10 GHz, and measure 20\% of the total power in each of the second order sidebands, with the carrier suppressed to less than 5\%. The remaining power is distributed among the first and higher order sidebands that do not interact with the atoms to drive any one or two photon transitions. We have modulated the EOM at half the hyperfine splitting of $^{87}$Rb to ensure that there is sufficient modulation depth to supress the carrier, as it otherwise forms a standing wave with the unmodulated beam that results in Bragg diffraction of the condensate and any outcoupled atoms. Finally, both beams are coupled into polarization maintaining fibers to produce a single spatial mode, and are collimated at the output to an $\sim$ 1 mm diameter. This results in approximately 4 and 8 mW of power after the fibers for the modulated and unmodulated beams respectively.

The beams are directed and focussed  onto the condensate as shown in Fig.\,\ref{AL}(c), using a 10 cm and 50 cm focal length lens for the horizontal and vertical beam respectively. This geometry was chosen to allow for polarization to be set such that a photon is absorbed from the $\omega_{+2}$ component in the horizontal beam ($\sigma^{+}$), and emitted into the unmodulated vertical beam ($\pi$), resulting in a momentum transfer of $\sqrt{2}\hbar k$ at 45$^\circ$ relative to gravity. If this is not done, the presence of a $\sigma^-$ component produces a second beam with the opposite momentum kick due to the lower sideband, $\omega_{-2}$.  It would be straightforward to remedy this if necessary by frequency filtering the modulated beam using an optical cavity. 

Based on our parameters, we estimate the maximum intensity at the condensate to be on the order of 10\,W/cm$^2$ leading to a maximum calculated Rabi frequency of $\Omega_{\textrm{\footnotesize{Rabi}}}^{(2)}=2\pi\times20$\,kHz. Our setup can be operated in the continuous weak-coupling regime, or in the pulsed strong-coupling regime. Following the work of Haine \textit{et al.}  \cite{Haine:2005aa,Haine:2006aa}, it would be straightforward to incorporate a squeezed optical beam as one of the Raman beams, allowing production of a quadrature squeezed atom laser.

\begin{figure}[ht]
\begin{center}
\includegraphics{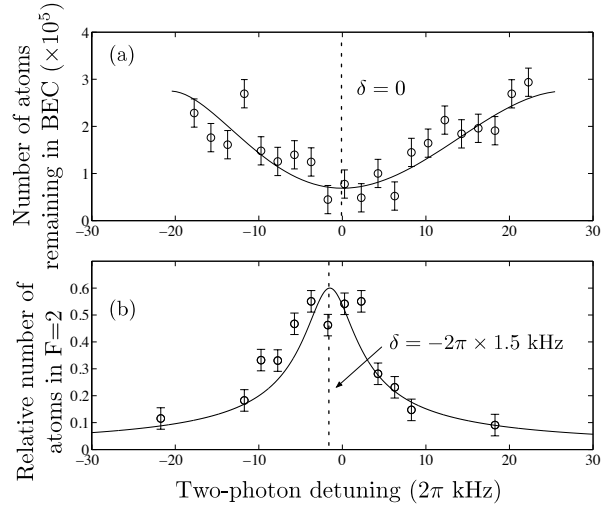}
\caption{\label{res} Resonance curves for the (a) weak- and (b) strong-coupling regimes, showing the lineshape of the resonance and the effect of the AC Stark shift. In (a), we use a 100 ms outcoupling pulse and plot the number of atoms remaining in the condensate as a function of two-photon detuning. The solid line is a fit to the experimental data using a Thomas-Fermi approximation, and integrating over a resonant slice of the wavefunction.  In (b), the relative number of atoms transferred to the $F=2$ ground state is plotted as a function of two-photon detuning. A Lorentzian envelope is fitted to the data (solid line) to determine the centre of the resonance. In both data sets, zero detuning corresponds to the centre of the Raman resonance for weak coupling, and error bars represent one standard deviation in the total atom number. }
\end{center}
\end{figure}

\section{Results and Discussion}

To determine the microwave frequency operating points, we measuring resonance curves for both weakly- and strongly-coupled systems. In the weakly-coupled system, we use a Rabi frequency of $\Omega_{\textrm{\footnotesize{Rabi}}}^{(2)} \simeq2\pi\times200$\,Hz, and a long Raman pulse of 100 ms, plotting the number of atoms remaining in the condensate in Fig.\,\ref{res}(a) as a function of the two-photon detuning, $\delta$. The solid line is a theoretical fit based on an integration of the Thomas-Fermi density, $|\psi(x,y,z)|^2=1/U(\mu-m/2(\omega_y^2y^2+\omega_r^2[x^2+z^2]))$ over a resonant slice of energy width. $\hbar \Omega_{\textrm{\footnotesize{Rabi}}}^{(2)}$, and assuming a flux directly proportional to this number times $\Omega_{\textrm{\footnotesize{Rabi}}}^{(2)}$ \cite{Robins:2006aa}. $U=4\pi\hbar^2a/m$ is the effective atomic interaction strength, $a$ the s-wave scattering length, $\mu=(15 N U \omega_y \omega_r^2 / 8\pi)^{2/5} (m/2)^{3/5}$ the chemical potential, with $N$ the total number of condensate atoms, and $m$ the atomic mass of $^{87}$Rb.  If we compare the microwave frequency corresponding to the minimum in this curve to the most recently published value for the ground state hyperfine splitting of $^{87}$Rb \cite{Bize:1999aa}, we find it to be in agreement to 1 part in $10^8$, where we have used a frequency counter locked to a GPS reference oscillator to calibrate our microwave frequency. $\delta=0$ is therefore equivalent to this point, taking account of the kinetic energy associated with the momentum transfer. In the strongly-coupled regime, a 60 $\mu s$ Raman pulse is used for our maximum measured Rabi frequency of $\Omega_{\textrm{\footnotesize{Rabi}}}^{(2)} \simeq 2\pi \times 10$\,kHz, and the relative number of atoms transferred to the $|2,0\rangle$ ground state is plotted as a function of $\delta$ in Fig.\,\ref{res}(b). The solid line is a Lorentzian fit to the data, based on the envelope of a two-state rate model, used only to extract the operating point from the data. This point occurs at $\delta=-2\pi\times1.5$\,kHz, and gives a direct measure of the maximum AC Stark shift in this system, in agreement with our previously stated estimate.

\begin{figure}[th]
\begin{center}
\includegraphics[scale=0.9]{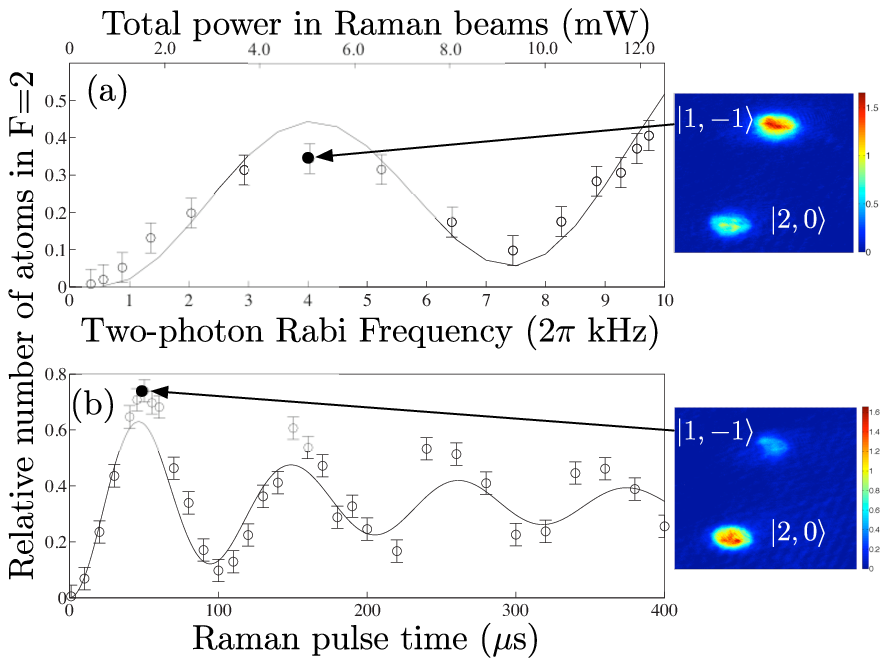}
\caption{\label{flop_calib} (a) Calibration of the two-photon Rabi frequency. The circles represent the relative number of atoms in the $|2,0\rangle$ state as a function of total power in the Raman beams. (b) Rabi oscillations between the $|1,-1\rangle$ and $|2,0\rangle$ hyperfine ground states of $^{87}$Rb. The relative number of atoms in the $F=2$ ground state is plotted as a function of pulse time, with a oscillation frequency of $\sim$ 10 kHz. Incomplete and decaying oscillations are observed due to an imperfect and decreasing overlap of the wavefunctions for the two coupled states in time. In both data sets, the error bars represent one standard deviation in total atom number and the solid line is a simulation of the GPE for our experimental parameters, which allows calibration of the Rabi frequency in (a).}
\end{center}
\end{figure}

Although we have estimated an upper bound on the maximum Rabi frequency of our outcoupler, accurate knowledge of the light intensity at the condensate is difficult to obtain. It is therefore important to definitively calibrate the Rabi frequency for any given setup. This is accomplished by comparing experimental data with a comprehensive 3D mean field theory. The Raman lasers are applied to the condensate for a fixed pulse length of 135 $\mu$s, and the power in the beams varied. This outcouples a pulse of atoms in the $|2,0\rangle$ state, whose relative number depends on $\Omega^{(2)}_{\textrm{\footnotesize{Rabi}}}$, and is plotted in Fig.\,\ref{flop_calib}(a) as a function of the total power in the two beams. The calibration is then extracted by fitting the experimental data to a 3D simulation of the Gross-Pitaevski equation (GPE) based on measured parameters of our system. Our maximum attainable Rabi frequency is on the order of $2\pi \times10$ kHz.

In Fig.\,\ref{flop_calib}(b), we demonstrate coherent Rabi oscillations of our two-level system at the maximum Rabi frequency for this setup. The solid line represent a 3D simulation of the GPE for our experimental parameters. For complete Rabi oscillations to occur, it is necessary for the spatial wavefunctions of each state to remain well overlapped over the duration of a Rabi cycle. As the $|2,0\rangle$ state  receives an initial momentum kick in addition to experiencing a net acceleration due to gravity, complete transfer was not observed, with a maximum transfer of $\sim$ 75\% in the first oscillation. Over time, and successive oscillations, the overlap continues to decrease with fewer atoms oscillating between the two states. This behaviour is qualitatively captured by our numerics.

We have also verified the ability to use our Raman laser system as a hyperfine state beamsplitter, by using a $\Delta m_F=0$ transition to transfer up to 95\% of a pulse atoms of freely falling in a $|1,0\rangle$ state to a $|2,0\rangle$ state; a crucial step towards a coherent Ramsey interferometer, and the subject of a future publication.

\section{Conclusion}
Our Raman laser system is a versatile and effective tool for coherently manipulating Bose-Einstein condensates. Operating via a pure two-state coupling, it produces atom lasers in a single atomic state and has the potential to transfer correlation statistics from a quadrature squeezed optical beam, making it a promising tool for investigating quadrature squeezing of atom lasers and entanglement in atomic beams. The same system can be operated as a coherent hyperfine state beamsplitter, a tool which is crucial for experiments in coherent Ramsey interferometry and precision measurement. Although we have demonstrated operation of this system using $^{87}$Rb, the optical setup is relatively straightforward compared with other phase coherent microwave systems, and can easily be transfered to other typical condensate species such as sodium, lithium, or cesium.


\begin{thebibliography}{10}
\bibitem{superradiance}
It should be noted that in previous work on Raman superradiance \cite{superrad1,superrad2}, atoms are removed from a condensate accompanied by a change in hyperfine ground state. Although it may be tempting to refer to this as an outcoupling process, we feel that this physical mechanism cannot be associated with the term \enquote{outcoupling} in the context of atom lasers. Firstly and foremost, the process transfers atoms to an untrapped state, which is unsuitable for atom lasers and their applications. Secondly, superradiance typically operates in a short pulse regime, and cannot be operated continuously.
\end{thebibliography}
\end{document}